\documentstyle[prl,aps,multicol,epsfig]{revtex}
\renewcommand{\d}{{\textrm d}}
\newcommand{\ai}{{\rm Ai}}
\newcommand{\bi}{{\rm Bi}}
\newcommand{\ee}{{\rm e}}
\newcommand{\ii}{{\rm i}}

\newcommand{\yt}{Y_t(x)}
\newcommand{\ytz}{Y_{t_0}(x)}
\newcommand{\ytt}{Y_T(x)}

\newcommand{\eqref}[1]{(\ref{#1})}
\newcommand{\yye}{{\mbox{\boldmath $y$}}}

\newcommand{\www}{{{\textbf w}}}
\newcommand\mean[1]{{\big<#1\big>}}

\begin{document}
\draft
\title{Dynamics of defect formation}
\author{Esteban Moro$^{1,2,*}$ and Grant Lythe$^{1,\dag}$}
\address{$^{1}$Center for Nonlinear Studies,\\
Los Alamos National Laboratory MS B285, New Mexico 87544 USA. 
\\ $^{2}$Departamento de Matem\'aticas and Grupo
Interdisciplinar de Sistemas Complicados,\\ Universidad Carlos III de 
Madrid, E-28911 Legan\'{e}s, Madrid, Spain} 
\date{\today}
\maketitle

\begin{abstract}
A dynamic symmetry-breaking transition with noise and inertia is analyzed. 
Exact solution of the linearized equation that describes the critical 
region allows precise calculation (exponent and prefactor) of the number 
of defects produced as a function of the rate of increase of the critical 
parameter. The procedure is valid in both the overdamped and underdamped 
limits. In one space dimension, we perform quantitative comparison with 
numerical simulations of the nonlinear nonautonomous stochastic partial 
differential equation 
 and report on signatures of underdamped dynamics.
\end{abstract}

\pacs{PACS number(s): 02.50Ey, 05.70Fh, 64.60-i}

\preprint{LANL}

\tighten

\begin{multicols}{2}
\narrowtext

 When a system that undergoes a symmetry-breaking transition is swept
through its critical point the initial symmetry is broken and domains are 
formed. Because of critical slowing down it is not possible to sweep 
adiabatically; the number of domains therefore depends on the rate of 
increase of the critical parameter. A new scenario for structure formation 
in the early universe and a proposal for its test in laboratory 
experiments resulted from the first understanding of the importance of 
this nonequilibrium effect 
\cite{zureknat1}.  Until recently, experimental \cite{exp} 
results tended to support the proposed scenario, but precise comparison 
 was not possible because neither experiment nor theory was
 confident of more than exponents. The situation is now changing,
with new experiments using quenches of liquid helium through the 
superfluid transition taking care to minimise vortex creation via flow 
processes \cite{new}. In this Letter we report new theoretical results: 
precise expressions for the number of defects and quantitative agreement 
with numerical results. 

 The phenomenon of a dynamic transition has been studied  
in the zero-dimensional case (pitchfork bifurcation) in the context of 
lasers \cite{eandm}--\cite{lille}. The time-dependence of the critical 
parameter produces a delay of the bifurcation given by 
\(\sqrt{2\mu|\ln\epsilon|}\) where \(\mu\) is the rate of increase of
the parameter and \(\epsilon\) the magnitude of additive fluctuations. 
Theoretical studies on spatially-extended systems 
 revealed a characteristic distance between kinks. The spatial
structure formed during the sweep through critical point from the 
symmetric to broken-symmetry regime is frozen in by the nonlinearity when, 
sufficiently far into the symmetry-broken regime, the system attains a 
metastable state 
\cite{grant1,zurekprl1,laguna,yates,fises}.
  Analytical progress is possible because the
critical region is well-described by an equation which, although 
stochastic and non-autonomous, is linear. Here we consider the influence 
of inertia; we derive the scalings and signatures of the overdamped and 
underdamped limits. 

The theory of dynamic transitions identifies three successive regimes in 
the evolution, as the critical parameter is increased. 
 In the earliest regime, 
 sufficiently far from the critical point, the evolution is
quasi-adiabatic: the ensemble of field configurations is a small 
perturbation of that found for constant parameters 
\cite{grant1,fises}. In the second region, close to the critical
point, the system can no longer react quickly enough to the 
time-dependence of the critical parameter 
\cite{zureknat1}.  Our treatment based on the
equation of motion, however, passes seamlessly between the first and 
second regions: in both, the field is everywhere small and precise 
calculation of the correlation function can be made from the linearized 
stochastic partial differential equation (SPDE) 
\cite{grant1,fises}.  We show that,
for the purposes of calculating the number of kinks formed, the end of the 
second, nonequilibrium, region is the key. In the final region, the 
spatial structure consists of narrow kinks separating long regions where 
the field is close to one of the minima of the potential. 
 The spatial structure is ``frozen in'' in the sense that the motion,
merging and occasional nucleation of kink-antikink pairs happens on a 
slower timescale than the process that formed them. The separation of 
timescales is especially marked at high damping and low temperature 
\cite{laguna}.

We shall consider the specific example of the stochastic process in one 
space dimension satisfying the following nonautonomous SPDE 
\cite{zurekprl1,yates,polaco}
\begin{eqnarray}
&\partial^2_t \yt -D\partial^2_x \yt + \gamma 
\partial_t \yt =\nonumber\\[3pt] &=g(t) \yt - \yt^3 +
 \epsilon\,\eta(x,t).
\label{eqGL}
\end{eqnarray}
The order parameter at time \(t\) and position \(x\), denoted by \(\yt\), 
is a real-valued random variable. The last term in 
\eqref{eqGL} is space-time noise, delta-function correlated in space
and time: \(\mean{\eta(x,t) \eta(x',t')} =\delta(x-x') \delta(t-t')\). 
 The fluctuation-dissipation relation
is enforced by setting \(\epsilon^2 =2\gamma k_{B}\Theta\), where 
\(\Theta\) is the temperature.

In our numerical simulations, the time-dependence of the critical 
parameter is: \(g(t)=\mu t\), starting at \mbox{\(g=-\tau < 0\)}. The initial 
conditions are 
\begin{equation}
\ytz =\partial_t\ytz =0, \qquad t_0=-\frac{\tau}{\mu} \label{ics}
\end{equation}
The simulations are performed on a domain \([0,L]\) that contains many 
kinks, 
 using periodic boundary conditions. Second order stochastic
time-stepping \cite{dynpf} was used for the spatially discretized version 
of \eqref{eqGL} \cite{grant1}. 

\begin{figure}
\vspace{0.1cm} 
\epsfig{file=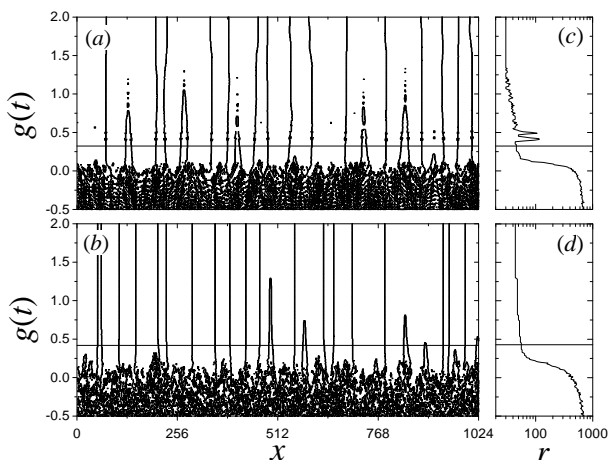,width=3.2in,clip=}
\caption{Space-time evolution. The positions of crossings of zero are
shown as a function of time for underdamped (a) and overdamped (b) 
dynamics. In (c) and (d) the corresponding numbers of crossings are shown. 
(Time increases upward.) The horizontal lines correspond to $g = \hat g$.} 
\label{stfig} 
\end{figure}

Typical time evolution is represented in Fig.\ \ref{stfig}, where each dot 
is the spacetime position of a zero crossing in one numerical realization. 
The system makes a transition from a regime with many zero crossings and 
typical values of the field close to zero, to a regime with few zeros, 
corresponding to the positions of kinks, separating large regions where 
\(\yt\) is close either to 
\(+\sqrt{g}\) or to \(-\sqrt{g}\). The transition takes place
at \(\hat g > 0\). For \(g>\hat g\), each zero of \(\yt\) corresponds to a 
well-defined kink or antikink. 

During the evolution preceding \(g=\hat g\), 
 the cubic term in \eqref{eqGL} is small and
 the linearized equation is a good approximation.
It is illuminating to nondimensionalize the linearized equation: 
\begin{eqnarray}
&\partial^2_{T} \ytt -\nu \partial^2_{x}\ytt +2\alpha \partial_{T} \ytt 
\nonumber\\ &=T \ytt + {\epsilon}{\mu^{-\frac12}}\eta(x,t),
\label{linspde} \cr
{\rm where}&\quad T=\mu^{\frac13}t,\quad 
 \alpha=\frac12\gamma\mu^{-\frac13},\quad
\nu=D\mu^{-\frac23}.
 \label{linspde2}
\end{eqnarray}
The dynamics can now be studied in terms of the characteristic time 
\(t=\mu^{-\frac13}\) and nondimensional damping \(\alpha\). The field 
\(\ytt\) satisfying \eqref{linspde} with initial conditions 
\eqref{ics} is Gaussian with mean zero at all times.
 The correlation function, 
\(c(x)=\mean{Y_{T}(x)Y_{T}(0)} \), changes its form
and amplitude with time. At any fixed time, there is the following 
relationship between \(c(x)\) 
 and the number of zeros: if \( c'(0)=0\) then
the mean number of zero crossings is a finite number given by 
\cite{ito,fises}
\begin{equation} 
\rho(T)=\frac{L}{\pi}\left(-\frac{c''(0)}{ c(0)}\right)^\frac12. 
\label{densz} 
\end{equation}

Analytical solution of \eqref{linspde} proceeds by separating into 
independent stochastic differential equations for each of the Fourier 
coefficients, \(\yye_T(k) 
 = L^{-\frac12}\int_0^L \ytt\ee^{\ii k\frac{2\pi}{L} x}\d x
\), whose time evolution is given by the SDE
\begin{eqnarray}
\partial^{2}_{T} \yye_T(k) + 2\alpha \partial_{T} \yye_T(k) =
\nonumber\\
=(T - \kappa^2) \yye_T(k)+ {(2\mu)^{-\frac12}}{\epsilon} \, \hat 
{\mbox{\boldmath $\eta$}}(T,k) 
\label{fmsde}
\end{eqnarray}
for integer \(k\) and where \( \kappa^2 =\nu\,k^2(\frac{2\pi}{L})^2\) and 
\(\mean{\hat {\mbox{\boldmath $\eta$}}(T,k) \hat {\bf \mbox{\boldmath
$\eta$}}(T',k')}=\delta_{k,k'}\delta(T-T') \). Each \(\yye_T(k)\) is 
Gaussian with mean zero \cite{exactsoln}. The variance grows exponentially 
fast for \(T-\kappa^2+\alpha^2 >1 \): 
\begin{eqnarray}
&\mean{\yye_T^*(k)\yye_T(k)} \to \pi\frac{\epsilon^2}{\mu} 
\Phi(T_0,\alpha,\kappa^2)\left(T-\kappa^2+\alpha^2\right)^{-\frac12}
\cr&\times \exp\left(\frac43(T-\kappa^2+\alpha^2)^{\frac32}
-2\alpha(T-\kappa^2)-\frac43\alpha^3\right), \label{ytvarapp} 
\end{eqnarray}
where 
\begin{eqnarray}
\Phi(T_0,\alpha,\kappa)=\ee^{\frac43\alpha^3}
\displaystyle\int_{T_0}^{\infty}\ai^2({S-\kappa^2+\alpha^2}) \ee^{2\alpha
{S}}\d {S}. 
\label{cdef}
\end{eqnarray}

No approximations have been made thus far in the solution of 
(\ref{linspde2}). We now consider the implications of the physical picture 
presented above for the relative values of the parameters. Firstly, for 
there to be a quasi-adiabatic first regime in the evolution, we require a 
sufficiently slow sweep: \(\mu \ll \tau^{\frac32}\) \cite{haberman}. 
Secondly, we require a well-defined value of the order parameter, \(g=\hat 
g\), marking the end of the second part of the evolution, implying 
\(\epsilon^2 \ll \mu\) \cite{dynpf}. 

 We adopt the following definition: 
\(\hat g =\mu^{\frac23}\hat T\) where
\(\hat T\) satisfies
\(
\mean{Y_{\hat T}^2(x)} =\delta\hat g
\).
(i.e., when \(\delta=1\), the first two terms on the RHS of \eqref{eqGL} 
are equally important.) We thus evaluate \(\hat g\) by solving 
\(\mean{Y_{\hat T}^2(x)}=\delta \hat g\) where
\begin{eqnarray}
\mean{Y_{T}^2(x)}
 =&\frac1L\sum_{k}\mean{\yye^*_T(k)\yye_T(k)}\cr
\simeq&(\lambda(T))^{-1}
(2\pi)^{-\frac12}\mean{\yye^*_T(0)\yye_T(0)} 
\label{yts}
\end{eqnarray}
and \(\lambda(T)=-2\nu\frac{\partial^2}{\partial \kappa^2} 
\ln\left(\mean{\yye^*_T(k)\yye_T(k)}\right)\).

The correlation function is the Fourier transform of 
\(\mean{\yye^*_T(k)\yye_T(k)}\). It emerges from the sweep past \(g=0\) 
with the form 
\(c(x)=c(0)\exp(-x^2/2\lambda^2(\hat T))\) \cite{grant1}.
 The number of zeros present at \(g=\hat g\) is thus
\begin{equation}
\rho=\frac1{\pi}\frac{L}{\lambda(\hat T)}.
\label{rlamb}
\end{equation}
 Our procedure is valid for arbitrary damping.
 We now examine the underdamped and the overdamped limits,
 defined by the parameter $\alpha = \frac{1}{2} \gamma
\mu^{-\frac13}$. The overdamped limit (studied in
\cite{grant1}) is recovered as $\alpha \to \infty$.

The {\em underdamped limit}. When \(\alpha\to 0\), \(\rho\) is only 
logarithmically dependent on \(\alpha\). In this limit \(\lambda(T) = 
2\nu^{\frac12}T^{\frac14}\), and the integral \eqref{cdef} has the 
asymptote 
\( \Phi(T_0,\alpha,\kappa^2)\to \Phi_1 |T_0-\kappa^2|^{\frac12} \)
 \cite{ctes}. As \(T\to \hat T\),
\begin{eqnarray}
\mean{Y_{T}^2(x)} \to
\pi \frac{\epsilon^2}{\mu} \Phi_1
\left(\frac{\tau}{4\pi D}\right)^{\frac12}
T^{-\frac34}\exp({\frac43 T^{\frac32}}), 
\label{myts}
\end{eqnarray}
and \(\hat g\) satisfies 
\begin{equation}
\hat g^{\frac32} =\frac34\mu\ln\left(
\frac{\mu}{\pi\epsilon^2}
(\frac{4\pi D}{\mu\tau})^{\frac12} 
\frac{\delta \hat g^{\frac74}}{\Phi_1}
 \right).
\label{ghatsat}
\end{equation}
Immediately before \(g=\hat g\), the number of zeros \eqref{densz} is a 
decreasing function of time, given by 
\begin{eqnarray}
\rho(T) =\displaystyle
\frac{L}{2\pi} \frac{T^{-\frac14}}{\nu^{\frac12}}.
 \label{ncerost}
\end{eqnarray}
The number of zeros present at \(\hat g\) for \(\alpha\to 0\) 
 is 
\begin{eqnarray}
\rho(\hat T) =\frac{L}{2 \pi}\frac{\mu^{\frac13}}{D^{\frac12}}
 {\left[ \frac34\ln\left(
\frac{\mu}{\pi\epsilon^2}
(\frac{4\pi D}{\mu\tau})^{\frac12} 
\frac{\delta\hat g^{\frac74}}{\Phi_1}
\right)
 \right]^{-\frac16}}.
 \label{nceros1}
\end{eqnarray}

In {\em the overdamped limit} the number of zeros is proportional to 
\(\mu^{\frac14}\) \cite{grant1,zurekprl1}.
  We show that the latter scaling is obtained in the
limit \(\alpha \to \infty\). Here \(\Phi(T_0,\alpha,\kappa) \to 
\Phi_2 \alpha^{-\frac12}\)
 \cite{ctes} and \(\lambda(T)^2 = 2\nu T/\alpha\). Thus
\begin{equation}
\mean{Y_{T}^2(x)} =
\pi\frac{\epsilon^2}{\mu}\frac{\Phi_2}{\alpha}
(4 \pi\nu T)^{-\frac12}\exp(\frac12\frac{T^2}{\alpha}) 
\label{mysod}
\end{equation}
and \(\hat g\) satisfies 
\begin{equation}
\hat g^2 =\mu\gamma\ln\left(
\epsilon^{-2}\gamma\delta(8 D\hat g^3)^{\frac12}
\right).
\label{ghatod}
\end{equation}
The number of zeros for overdamped slow passage is 
\begin{eqnarray}
\rho(\hat T)
 =& \displaystyle
\frac{L}{2 \pi}\frac{(\mu\gamma)^{1/4}}{D^{\frac12}}
\left[\ln(
\epsilon^{-2}\gamma\delta(8D\hat g^3)^{\frac12}
)\right]^{-\frac14}. 
 \label{nceros2}
\end{eqnarray}

\begin{figure}
\hspace{-0.6cm} 
\epsfig{file=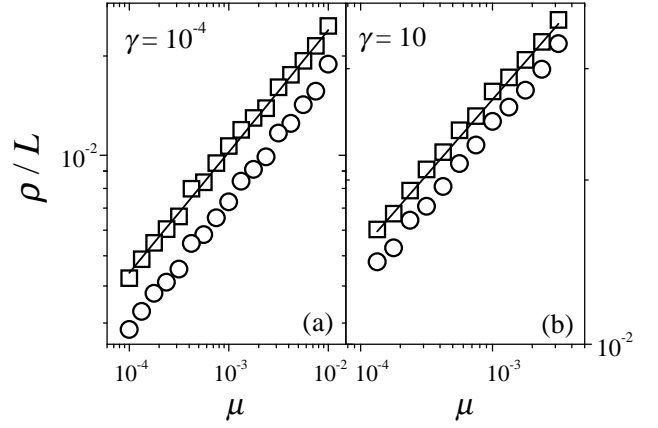,width=3.5in,clip=}
\caption{Density of zeros for the underdamped (a) 
and overdamped (b) cases. The lines are the analytical approximations 
\eqref{nceros1} and \eqref{nceros2}.
Averages from numerical simulations of \eqref{eqGL} are shown at 
 $g =\hat g$ (squares) and  $g =\tau$ (circles),
with errors of symbol size or smaller. ($D=\tau=1$, $\Theta=5\times 
10^{-9}$.)} 
\label{fig2}
\end{figure}

Equations \eqref{nceros1} and \eqref{nceros2} are the main results of this 
Letter: the number of created defects scales with the sweep rate as $\rho 
\sim \mu^{\frac14}$ for the underdamped case and $\rho 
\sim \mu^{\frac13}$ for the overdamped regime.  We have performed
extensive quantitative comparison with numerical simulations of 
\eqref{eqGL}. Two examples are shown in Fig.\ \ref{fig2}. Our
analytical predictions at instant $g = \hat g$ are very accurate. Although 
the exponents can be obtained from dimensional analysis 
\cite{zureknat1,laguna}, logarithmic corrections produce
small deviations in numerically-estimated exponents at finite damping. 
  No evidence has been found for
the region of \(\rho \sim \mu^{\frac12}\) scaling, predicted in 
\cite{polaco} from an approximation that replaced \eqref{fmsde} by a
first-order equation. 

Because the number of defects at \(\hat g\) is typically much larger than 
the equilibrium density at temperature \(\Theta\) \cite{further}, their 
number decreases after \(\hat g\) as kink-antikink pairs annihilate (see 
Fig.\ \ref{stfig}). The smaller the damping, the more rapidly this 
annihilation proceeds \cite{laguna}. In Fig.\ \ref{fig2} we have also 
plotted the number of zeros at \(g=\tau\). While the number of zeros is 
reduced, the scaling with \(\mu\) seen at $g = \hat g$ is preserved. 

\begin{figure} \vspace{0.1cm} 
\epsfig{file=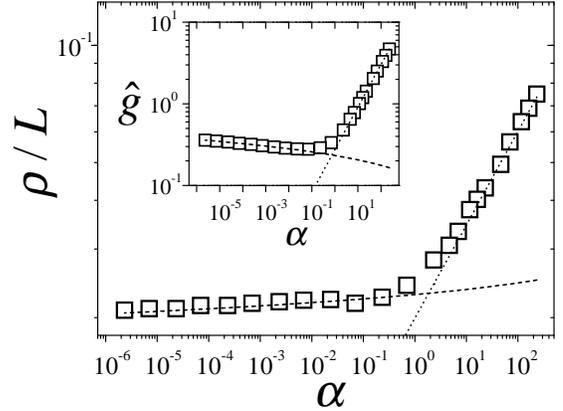,width=3.2in,clip=} 
\caption{From the underdamped to the overdamped regime:
 density of zeros vs $\alpha$ for fixed $\mu$. 
Squares are numerical averages. The main plot shows the density of defects 
at the instant $\hat g$ and the analytical predictions for the underdamped 
\eqref{nceros1} and the overdamped \eqref{nceros2} regimes. We
measured \(\hat g\) numerically as the instant when $\mean{Y^2_{\hat 
T}(x)}= \frac{1}{2} \hat g$. In the inset we plot the value thus obtained 
vs\ (\ref{ghatsat}) and (\ref{ghatod}). ($\mu = 10^{-2}$, $D = \tau = 1, 
\Theta = 5\times 10^{-9}$.)} \label{fig2a} 
\end{figure}

The crossover between regimes, represented in Fig.~\ref{fig2a}, takes 
place when the nondimensional damping \mbox{\(\alpha = 
\frac{1}{2}\gamma\mu^{-\frac13} \simeq 1\)}.  At small damping the
dependence of the number of defects on damping is only logarithmic, 
(\(\epsilon^2\propto \gamma\)). At large damping, $\rho \sim 
\gamma^{\frac14}$. 

\begin{figure} \vspace{0.1cm} 
\epsfig{file=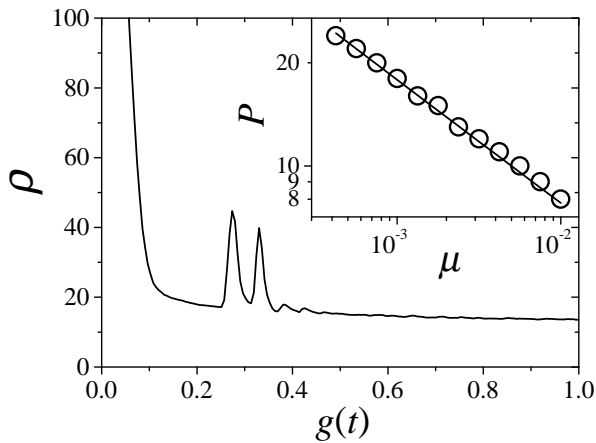,width=3.4in,clip=} 
\caption{``Bounce back'': The number of zeros oscillates 
after \(g=\hat g\) for small damping. The graph of \(r\) vs time is for 
$\mu=4\times10^{-3}$, $\gamma = 10^{-4}$, $\epsilon = 10^{-6}$ and $D = 
\tau = 1$. The inset plots the period of the oscillations 
 obtained at different values of \(\mu\), with
\(\gamma = 10^{-4}\).  The straight line is 
 $P = 2\pi / \sqrt{2 \hat g}$
 with $\hat g$ given by \ (\ref{ghatsat}).}
\label{fig3} 
\end{figure}

Apart from the scaling of the number of defects with \(\mu\), a different 
signature of underdamped dynamics can be seen in in Fig.\ \ref{stfig} and 
Fig.\ \ref{fig3}: multiple ``bounce back'' of the number of zeros soon 
after \(g=\hat g\). The phenomenon has been reported in simulations of a 
sudden quench (\(\mu=\infty\)) 
\cite{luis}.  We propose the following
interpretation. In a dynamic transition at low damping, domains reach a 
minimum of a potential well with a finite velocity and therefore oscillate 
about it for a time. Parts of some domains 
 recross the crest of the instantaneous potential barrier
during these oscillations. This yields an estimate of 
 the frequency of the oscillations: $\sqrt{2 \hat g}$,
corresponding to harmonic oscillations about the minimum. 
  In our simulations, two well-defined bumps are
typically seen in the number of defects vs time. From this we are able to 
measure the period, \(P\) of the oscillations in the number of zeros; 
despite the nonlinearity, it is very well approximated by $P = 2 \pi 
/\sqrt{2 \hat g}$ (See Fig.\ \ref{fig3}). 

The procedure carried out in this Letter for the real 
 equation \eqref{eqGL} can be applied to other equations exhibiting
continuous transitions \cite{grant1} and in more than one space dimension 
\cite{grant1,yates}. The scalings are not sensitive to the particular 
equations chosen, but they are sensitive to any breaking of the exact 
\(Y\to -Y\) symmetry in the equation of motion. 

In summary, we derive quantitative predictions for the number of defects 
formed in a symmetry-breaking transition in one space dimension by 
analyzing the dynamics in the critical region, where the system is out of 
equilibrium regardless of how slowly the critical parameter is changed. 
Prefactors are calculated, so no fitting necessary. Underdamped slow 
passage results in a defect density proportional to \(\mu^{\frac13}\) and 
produces characteristic oscillations in the number of zeros. Experiments 
where liquid Helium is expanded through the Lambda Transition re 
now reaching the point where quantitative comparisons can be made.

We are grateful for Angel Sanchez's comments on the manuscript. E. Moro 
thanks the CNLS for its hospitality.

\end{multicols}


\begin{references}
\bibitem{exactsoln}
The solution of \eqref{fmsde} is 
\(\yye_T(k) = {\pi}{\epsilon}{(2\mu)^{-\frac12}}\Big[\)
\begin{eqnarray}
\displaystyle\int_{t_0}^{t}\ee^{-\alpha(T-S)}
\Big(\ai&(T-\kappa^2+\alpha^2)\bi(S-\kappa^2+\alpha^2)&\cr
-\bi&(T-\kappa^2+\alpha^2)\ai(S-\kappa^2+\alpha^2)& 
\Big)\d\www_S(k)\Big];
\nonumber
\end{eqnarray}                       
$\www_S(k)$ is a Wiener process, $\ai$ and $\bi$ are Airy functions.
\bibitem{ctes}
The constants used in the calculations are \filbreak 
$\Phi_1=\displaystyle\lim_{T_0 \rightarrow -\infty} 
|T_0|^{-\frac12}\int_{T_0}^{\infty} \ai^2(S) \d S =0.318\ldots$ 
 and \filbreak
$\Phi_2=\displaystyle\lim_{\alpha \rightarrow \infty}\alpha^{\frac12} 
\ee^{\frac43\alpha^3}\int_{-\infty}^{\infty} \ai^2(S+\alpha^2)
\ee^{2\alpha S} \d S =(8\pi)^{-\frac12}$.
\end{references}
\end{document}